%Paper: cond-mat/9302005
%From: hans <HTC004%DJUKFA11.BITNET@vm.cnuce.cnr.it>
%Date: Tue, 02 Feb 93 09:41:02 +0100

%%%%%%%%%%%%%%%%%%%%%%%%%%%%%%%%%%%%%%%%%%%
%   DENSITY PATTERNS IN GRANULAR MEDIA    %
%                                         %
%  Gerald H. Ristow and Hans J. Herrmann  %
%%%%%%%%%%%%%%%%%%%%%%%%%%%%%%%%%%%%%%%%%%%
%
%\magnification=1200
\baselineskip=16pt
\normallineskip=8pt
\vsize=23 true cm
\hsize=15 true cm
\overfullrule=0pt
\voffset=-1.0 true cm
\pageno=1
\font\bigfont=cmr10 scaled\magstep1
\footline={\hss\tenrm\folio\hss}
\def\ref#1{$^{[#1]}$}
%------------------------ paper ---------------------
\vskip 2.0truecm
\rightline{HLRZ 2/93}
\bigskip\bigskip
\centerline{\bf \bigfont DENSITY PATTERNS IN GRANULAR MEDIA}
\bigskip\bigskip\bigskip\bigskip\smallskip
\centerline{\bf Gerald H.~Ristow$^*$ and Hans J.~Herrmann}
\medskip
\centerline{H\"ochstleistungsrechenzentrum, KFA,}
\centerline{D-5170 J\"ulich, Germany,}
\bigskip\bigskip\bigskip\vskip 2.5truecm
\noindent{\bf Abstract} \par
\noindent Recent experiments by Baxter et al.
showed the existence of density waves in granular material flowing out of
a hopper. We show, using Molecular Dynamics Simulations,
that this effect is a consequence of static
friction and find that these density fluctuations
follow a $1/f$ spectrum. The effect is enhanced when the opening
angle of the hopper decreases.
\vfill\eject
%------------------------------------------------------------------
\bigskip
Moving dry granular media, like sand, show a rich variety of rather
astonishing and scarcely understood phenomena\ref{1-2}.  Famous are the
so-called ``Brazil nut'' segregation\ref{3-5} and the heap formations that
occur under vibrations\ref{6-8}.  More recently a series of experiments
have given evidence that under certain circumstances density patterns are
generated inside the flowing medium. Baxter et al.\ref{9}, for instance,
visualized wave-like patterns emanating from the outlet of a two
dimensional wedge-shaped hopper using X-rays.  Also previous authors\ref
{10-12} had noted the formation of similar structures. Similarly rather
erratic shock-like density waves have been observed in flow through
pipes\ref {13} and down inclined planes\ref{14}.  Another experimentally
observed ubiquous phenomenon in granular media seems to be $1/f^\alpha$
noise. Lui and Jaeger\ref{15} recently measured the acceleration of a
particle inside a bulk of glass beads that where excited by a small
amplitude vibration. Its Fourier spectrum in time showed power law decay
over many orders of magnitude.  Baxter\ref{16} also observed power law
decay in the frequency dependent forces that act on the wall of a hopper.
For avalanches going down the slope of a sand pile theoretical
considerations of self-organized criticality\ref{17} led to the proposal
that their size and life times distributions were power laws which was in
fact only verified experimentally on very small piles\ref{18}.

The existence of these erratic density and force inhomogeneities are
intimately related to the ability of granular materials to form a hybrid
state between a fluid and a solid: When the density exceeds a certain
value, the critical dilatancy\ref{19,20}, granular
materials are resistant to
shear, like solids, while below this density they will ``fluidify''.  In
the presence of density fluctuations the rheology therefore can become
rather complex. Two microscopic facts seem to be responsible for the
strong density fluctuations: On one hand one has in granular media solid
friction between the grains.  This means that when particles are pushed
against each other a finite force is needed to start or maintain a
relative tangential motion between them. On the other hand a granular
material is internally disordered giving a natural source of noise. The
strong non-linearity coming from friction produces instabilities in the
density which enhance the fluctuations coming from the noise.

Various attempts have been made to formalize and quantify the complicated
rheology of granular media.  Continuum equations of motion \ref{21}, a
cellular automaton\ref{22} and a random walk approach\ref{23} have been
proposed. But none of them has yet been able to explain these
heterogeneous waves.  This is why we chose to study these phenomena using
Molecular Dynamics (MD) simulations of inelastic particles with static and
dynamic friction in two dimensional systems.  In fact, MD
simulations\ref{24,25} have already been applied to granular media to
model segregation\ref{5}, outflow from a hopper\ref{26,27}, shear
flow\ref{28}, convection cells on vibrating plates\ref{29,30}, avalanches
on a sand pile\ref{31}, flow through a pipe\ref{32} and others.

We consider a system of $N$ spherical particles
of equal density and with diameters $d$ either all equal or
chosen randomly from a Gaussian distribution of width $w$ around
$d_0 = 1$~mm. These particles are placed into a hopper having an
opening angle $\theta$ and at the bottom an opening of diameter $D$.
When two particles $i$ and $j$ overlap (i.e. when their
distance is smaller than the sum of their
radii) three forces act on particle $i$: 1.) an elastic restoration force
$$\vec f^{(i)}_{el} = Y ( \mid \vec r_{ij} \mid - {1 \over 2}(d_i
+ d_j)) {\vec r_{ij} \over \mid \vec r_{ij} \mid }\ \ ,\eqno(1a)$$
where $Y$ is the Young modulus and $\vec r_{ij}$ points from particle
$i$ to $j$;
2.) a dissipation due to the inelasticity of the collision
$$\vec f^{(i)}_{diss} = -\gamma m_{eff} (\vec v_{ij}\cdot \vec r_{ij})
{\vec r_{ij} \over \mid \vec r_{ij} \mid ^2}
= -\gamma m_{eff} v^n_{ij}\ \ ,\eqno(1b)$$
where $\gamma $ is a phenomenological dissipation coefficient and
$\vec v_{ij} = \vec v_i - \vec v_j$ the relative velocity between the
particles;
3.) a shear friction force which in its simplest from can be
chosen as
$$\vec f^{(i)}_{shear} = -\gamma_s m_{eff} (\vec v_{ij}\cdot \vec t_{ij})
{\vec t_{ij} \over \mid \vec r_{ij} \mid ^2}
= -\gamma_s m_{eff} v^t_{ij}\ \ ,\eqno(2a)$$
where $\gamma_s$ is the shear friction coefficient and
$\vec t_{ij} = (-r^y_{ij}, r^x_{ij})$ is the vector
$\vec r_{ij}$ rotated by 90$^\circ $.
Eq.~2a is a rather simplistic description of shear friction. In
our case it is important to include real static
friction which is done by a static
friction force\ref{33}:
When two particles start to touch each other, one puts a ``virtual''
spring between the contact points of the two particles.
Be $\delta s$ the {\it total} shear displacement of this spring
during the contact and $k_s \delta s$ the restoring
frictional force (static friction). The maximum value
of the restoring force is then according to Coulomb's criterion
proportional to the normal force $F_n$ and the proportionality
constant is the friction coefficient $\mu$. Cast into a formula this
gives a friction force
$$\vec f^{(i)}_{friction} = -{\rm sign}(\delta s) {\rm min}(k_s
m_{eff} \delta s, \mu F_n)\ \ \ \ .\eqno(2b)$$
where $\delta s$ is the shear displacement integrated over
the entire collision time. When particles are no longer
in contact with each other the spring is removed.
It is, however, not straightforward to implement the above
technique when the particles are allowed to rotate, i.e. to be able
to roll on each other. Therefore we did not take angular
momenta into account.
In fact, when particles have strong deviations from the spherical
shape rotations are strongly suppressed.
Often it is however useful to go an intermediate way and to
include dynamic friction but not static friction and allow
for the particles to have rotations\ref{5,27,29}. In that
case one uses a combination of eqs.~2a and 2b:
$$\vec f^{(i)}_{dyn} = -{\rm min}(\gamma_s m_{eff} v^t_{ij},
\mu F_n)\eqno(2c)$$
and introduces also equations of motion for the angular momentum
of the particles.

When a particle collides with a wall the same forces act as if it would
have encountered another particle of diameter $d_0$ with infinite mass at
the collision point.  The walls are in fact made out of small particles
themselves and in order to introduce roughness on the wall these particles
are chosen randomly from a distribution of two radii. The only external
force acting on the system is gravity $g \approx -10$m/s$^2$.

As initial positions of the particles we considered that they are placed
at random positions inside a space several times as high as the dense
packing. The initial velocities are set to zero.  After that the particles
are allowed to fall freely under gravity.  We took a Young modulus of
$Y=10^6$g/s$^2$ and a time step of
$\Delta t = 3.3\cdot 10^{-6}$s and run our
program on 8 or 16 processors of an Intel iPSC/860 and an IBM
RS/6000--550.

In Fig.~1 we see a snapshot of the outflowing particles at four different
time steps for material parameters consistent with
the experiment of Baxter et
al.\ref{2}. We clearly see that close to the outlet large holes appear
which then propagate in an attenuated form upwards.  These patterns
quickly vary in time. When the opening angle $\theta$ of the hopper is
reduced the contrast in the patterns becomes more pronounced. The holes
are in general stretched in the horizontal direction but their shape seems
rather random. Particularly striking are these structures when watched in
a movie.  When the static friction $\mu$ is switched off the structures
disappear and the density of the outflowing particles becomes homogeneous.
This agrees with Baxter et al.'s experimental observation that density
patterns only occur for rough and not for smooth sand. When the friction
with walls is switched off the density waves also disappear.

A more quantitive approach can be made by measuring the local densities
$\rho$. We binned space in units of 1.56$d_0$ and counted the number of
particles that have their center of mass inside the box averaged over 100
consecutive iteration steps.  This density is plotted in Fig.~2 as a
function of time and space. We see that the low density regions form
curved stripes pointing to rather short lived waves. No regular structure
in the distance and magnitude of the waves can be noted.  They rather look
like independent shock waves with random amplitudes coming in a random
sequence.  Larger waves usually have small densely packed precursors.
Strong similarities can in fact be seen with space-time plots of the
density of granular media in pipes\ref{32} and
of cars on highways having traffic jams\ref{34}.

In Fig.~3a we see the density profile as a function of time for a point
four particle diameters above the outlet. No regularity can be seen. The
over four components averaged Fourier transformed data are shown in a
log-log plot in Fig.~3b. Clearly they fall on a straight line over nearly
two decades. The slope is about -1.  This means that we have found $1/f$
noise. This is consistent with the picture that has emerged from the
mentioned experiments.

When particles of equal size are taken we observed equally well developed
density patterns and find roughly the same power law decay of the
spectrum.  The effect is reduced when the diameter $D$ of the outlet
becomes too large. If it is too small the flow of sand can entirely stop
due to arching. The critical diameter $D_0$ when this arching sets in has
been studied before with similar techniques\ref{26} where it was found
that $D_0$ is larger when the particles have the same size. When we consider
smooth walls, i.e. all wall particles
having the same radii, we do not find density waves and the power
spectrum looks significantly different. It shows
an upwards curved slope with increasing frequency which one also finds when
configurations block during the outflow. A similar effect was also
found in simulations of flow on an inclined plane\ref{35}.

Baxter et al. also observed that there exists a critical angle $\theta_0$
above which stagnation regions exist next to the walls of the container.
We also found these regions one example is shown in Fig.~4.

We have observed in a very simple modelization that density waves are
generated and distributed with frequency like $1/f$. Two ingredients were
found essential to generate them: static friction and disorder.  The
static friction tends to align the particles, i.e. to form fronts of
particles moving exactly with the same vertical velocity. These fronts are
nucleated randomly at the walls. Their size
distribution (density contrast) comes by itself into a critical state
namely a power law distribution. It therefore has the properties of
self-organized criticality (SOC)\ref{17}.  It is, however, very important
to notice that our simulations were made for rather small systems as
compared to real systems. It could therefore be that for systems of
millions of particles a cut off exists in this power law.

We have shown in this paper that similar to the avalanches that one
observes on the surface of a sand pile also inside the bulk of granular
material one has avalanche behavior which like the ones on the surface
shows self-organized criticality on small scales\ref{18}. Surface
avalanches for large piles however seemed to have a characteristic size
due to inertia\ref{36}. In fact, rather
avalanches in the holes than in the mattered substance
could be relevant here and one might speculate that the
bulk avalanches might be a better example for asymptotic SOC than the ones
on the surface. The mechanisms that generate the patterns are similar but
not identical to the original sandpile models. While the static friction
similarly generates waiting times with a threshold it is not the motion of
the sand itself that constitutes the avalanches but it is the group
velocity of the holes between them: An individual particle can easily go
from one dense region to the other by flying fast through a region of low
density. There is therefore a backflow of information similar to the
jamming on highways\ref{34}.

Although our simulations are two dimensional, we think that they do
capture the essential mechanisms that occur also in three dimensional
experiments. It should be mentioned that in fluidized beds (low Bagnolds
number) where the granular medium is surrounded by a fluid and the
hydrodynamic interactions become important a similar phenomenon as the one
described here, called slugging, is observed\ref{37}. The mechanisms
involved seem, however, quite different.
\vskip2cm
$^*$ present address: Groupe de Mati\`ere Condens\'ee et Materiaux,
     Universit\'e de Rennes I, Campus de Beaulieu, 35000 Rennes, France
\vskip1cm
\noindent {\bf References}
\item{1.} H.M. Jaeger and S.R. Nagel, Science {\bf 255}, 1523 (1992)
\item{2.} S.B.Savage, in
{\it Disorder and Granular Media} ed. D. Bideau
(North-Holland, Amsterdam, 1992);
S.B. Savage, Adv. Appl. Mech., {\bf 24}, 289 (1984);
C.S. Campbell, Annu. Rev. Fluid Mech. {\bf 22}, 57 (1990)
\item{3.} J.C. Williams, Powder Techn. {\bf 15}, 245 (1976)
\item{4.} A.~Rosato, K.J.~Strandburg, F.~Prinz and R.H.~Swendsen
     Phys.Rev.Lett.~{\bf 58}, 1038 (1987) and Powder Techn. {\bf49},
     59 (1986); P. Devillard, J. Physique {\bf 51}, 369 (1990)
\item{5.} P.K. Haff and   B.T. Werner, Powder Techn.
     {\bf 48}, 239 (1986)
\item{6.} M. Faraday, Phil. Trans. R. Soc. London {\bf 52}, 299 (1831)
\item{7.} P.~Evesque and J.~Rajchenbach, Phys.~Rev.~Lett.~{\bf62},
     44 (1989); C.~R. Acad. Sci. Ser.~2, {\bf 307}, 1 (1988) and
     {\bf307}, 223 (1988);
     C. Laroche, S. Douady and S. Fauve, J. de Physique
     {\bf 50}, 699 (1989);
     P. Evesque, J. Physique {\bf 51}, 697 (1990);
      J. Rajchenbach, Europhys. Lett. {\bf 16}, 149 (1991)
\item{8.} J. Walker, Sci. Am. {\bf 247}, 167 (1982);
     F. Dinkelacker, A. H\" ubler and E. L\" uscher, Biol. Cybern.
     {\bf 56}, 51 (1987)
\item{9.} G.W.~Baxter, R.P.~Behringer, T. Fagert and G.A.~Johnson,
     Phys.~Rev.~Lett. {\bf 62}, 2825 (1989)
\item{10.} O. Cutress and R.F. Pulfer, Powder Techn. {\bf 1}, 213 (1967)
\item{11.} A. Drescher, T.W. Cousens and P.L. Bransby, Geotechn.
{\bf 28}, 27 (1978)
\item{12.} R.L. Michalowski, Powder Techn. {\bf 39}, 29 (1984);
     J.D. Athey, J.O. Cutress and RF. Pulfer Chem. Eng. Sci. {\bf 21}, 835
(66);
    P.M. Blair-Fish and P.L. Bransby, J. Eng. for Industry {\bf 95}, 17 (73);
   J. Lee, S.C. Cowin and J.S. Templeton, Trans. Soc. Rheol. {\bf 18}, 247 (74)
\item{13.} T. P\"oschel, preprint HLRZ ../92
\item{14.} D. Bideau, private communication
\item{15.} C.-h. Lui and S.R. Jaeger, Phys. Rev. Lett. {\bf 68},
 2301 (1992) \& preprints
\item{16.} G.W. Baxter, PhD thesis
\item{17.} P. Bak, Tang and Wiesenfeld, Phys. Rev. Lett. {\bf 59},
381 (1987)
\item{18.} G.A. Held, D.H. Solina, D.T. Keane, W.J. Horn and
 G. Grinstein, Phys. Rev. Lett. {\bf 65}, 1120 (1990)
\item{19.} O. Reynolds, Phil. Mag. S. {\bf 20}, 469 (1885)
\item{20.} Y.M. Bashir and J.D. Goddard, J. Rheol. {\bf 35}, 849 (1991)
\item{21.} S.B. Savage, J. Fluid Mech. {\bf 92}, 53 (1979);
     G.M. Homsy, R. Jackson and J.R. Grace, J. Fluid Mech.
     {\bf 236}, 477 (1992);
S.B. Savage and K. Hutter, J. Fluid Mech. {\bf 199}, 177 (1989)
\item{22.} G.W. Baxter and R.P. Behringer, Phys. Rev. A {\bf 42},
    1017 (1990), Physica D {\bf 51}, 465 (1991)
\item{23.} H.~Caram and D.C.~Hong, Phys.~Rev.~Lett.~{\bf67}, 828
    (1991)
\item{24.} M.P.~Allen and D.J.~Tildesley, {\it Computer Simulation
     of Liquids\/}, Oxford University Press, Oxford, 1987
\item{25.} D.~Tildesley, in {\it Computational Physics\/}, edited by
      R.D.~Kenway and G.S.~Pawley, NATO Advanced Study Institute,
      Edinburgh University Press, 1987
\item{26.} G. Ristow, J. Physique I {\bf 2}, 649 (1992) and I.J.M.P.C.
\item{27.} D.C. Hong and J.A. McLennan, Physica A {\bf 187}, 159 (1992)
\item{28.} C.S. Campbell and C.E. Brennen, J. Fluid Mech.
    {\bf 151}, 167 (1985); P.A. Thompson and G.S. Grest,
    Phys. Rev. Lett. {\bf 67}, 1751 (1991);
    D.M. Hanes and D.L. Inman, J. Fluid Mech. {\bf 150}, 357 (1985);
    O.R. Walton and R.L. Braun, J. Rheol. {\bf 30}, 949 (1986)
\item{29.} J.A.C. Gallas, H.J. Herrmann and S. Soko\l owski,
      Phys. Rev. Lett., {\bf 69}, 1371 (1992)
\item{30.} Y-h. Tagushi, Phys. Rev. Lett. {\bf 69}, 1367 (1992)
\item{31.} J. Lee and H.J. Herrmann, preprint HLRZ 44/92
\item{32.} T. P\"oschel, preprint HLRZ 47/92
\item{33.} P.A. Cundall and O.D.L. Strack,
       G\'eotechnique {\bf 29}, 47 (1979)
\item{34.} K. Nagel and M. Schreckenberg, preprint
\item{35.} T. P\"oschel, preprint HLRZ 18/92
\item{36.} H.M. Jaeger, C.-h. Lui and S. Nagel, Phys. Rev. Lett.
      {\bf 62}, 40 (1988); P. Evesque, Phys. Rev. A {\bf 43}, 2720 (1991)
\item{37.} J.F. Davidson and D. Harrison, {\it Fluidization},
      (London Academic, 1971)
%\item{16.} J.A.C. Gallas, H.J. Herrmann and S. Sokolowski,
%      Physica A, in press
%\item{18.} P. Evesque, E. Szmatula and J.-P. Denis, Europhys.
%    Lett. {\bf 12}, 623 (1990); O. Zik and Stavans, Europhys.
%    Lett. {\bf 16}, 255 (1991); O. Zik, J. Stavans and
%    Y. Rabin, Europhys. Lett. {\bf 17}, 315 (1992)
%item{22.} J.A.C. Gallas, H.J. Herrmann and S. Sokolowski,
%     J. Physique II, {\bf 2}, 1389 (1992)
%\item{31.} H. Schmidt and I. Peschl, F\"ordern u. Heben
%{\bf 15}, 606 (1965)
%\item{34.} R.A. Bagnold, Proc. Roy. Soc. London A {\bf 295}, 219 (1966)
%H.M. Jaeger, C.-h. Liu, S.R. Nagel and T.A. Witten,
%Europhys. Lett. {\bf 11}, 619 (1990)
\bigskip
\vfill\eject
\noindent{\bf Figure Captions}
\medskip
\item{\bf Figure 1.} In this sequence, we show the outflow behavior with
     static friction for an initial configuration of roughly 1500 particles.
     The parameters were chosen as following: $\gamma=500$Hz, $\gamma_s=250$Hz,
     $k_s=1000$g/s$^2, \theta=30^\circ, D= 8 d_0, \mu=0.5$.
     (a) initial configuration, (b) after 10,000 iterations,
     (c) after 100,000 iterations, (d) after 140,000 iterations
\item{\bf Figure 2.} This figure shows the spatial
     density fluctuations (vertical axis)
     for a typical outflow simulation as a function of time
     (horizontal axis) for $\gamma=100$Hz, $\gamma_s=500$Hz, $k_s=1000$g/s$^2,
     \theta=10^\circ, D=10 d_0$ and $\mu=0.5$.
     Dark areas mark regions with lower densities.
\item{\bf Figure 3.} We consider an outflow sequence for the same parameter
     values as in figure 1 but we constantly fill in particles from above.
     a) shows the density fluctuations 6$d_0$ above the hole.
     One clearly sees a quite random behavior.
     b) shows the log-log plot of the FFT analysis of the density fluctuations.
     A straight line with slope -1 is drawn to guide the eye.
\item{\bf Figure 4.} Due to static friction, 314 particles remain in the hopper
     for an opening angle of $\theta=150^\circ$ ($\gamma=100$Hz,
     $\gamma_s= 0$Hz, $k_s=1000$g/s$^2, D=10 d_0$ and $\mu=0.5$).

\bye